# Estimation of genetic diversity in viral populations from next generation sequencing data with extremely deep coverage

Jean P. Zukurov[1,*], Sieberth do Nascimento-Brito[2,5,*], Angela C. Volpini[6], Guilherme C. Oliveira[6], Luiz Mario R. Janini[1,2] and Fernando Antoneli[3,4,§]

[1]Departmento de Medicina, EPM-UNIFESP

[2]Departamento de Microbiologia, Imunologia e Parasitologia, EPM-UNIFESP

[3]Departmento de Informática em Saúde, EPM-UNIFESP

[4]Laboratório de de Biocomplexidade e Genômica Evolutiva, EPM-UNIFESP

Escola Paulista de Medicina (EPM), Universidade Federal de São Paulo (UNIFESP), São Paulo, Brazil.

[5]Departamento de Microbiologia e Imunologia Veterinária, Universidade Federal Rural do Rio de Janeiro (UFRRJ), Rio de Janeiro, Brazil.

[6]Genomics and Computational Biology Group, Centro de Pesquisas René Rachou (CPqRR), Fundação Oswaldo Cruz (FIOCRUZ), Belo Horizonte, Brazil

E-mail addresses:

JPZ: zukurov@outlook.com

SNB: sieberth@ufrrj.br

ACV: avolpini@cpqrr.fiocruz.br

GCO: oliveira@cpqrr.fiocruz.br

LMRJ: janini@unifesp.br

FA: fernando.antoneli@unifesp.br

[§]Corresponding author fernando.antoneli@unifesp.br

*These authors have contributed equally to this work.

**Abstract**

**Background.** In this paper we propose a method and discuss its computational implementation as an integrated tool for the analysis of viral genetic diversity on data generated by high-throughput sequencing. The main motivation for this work is to better understand the genetic diversity of viruses with high rates of nucleotide substitution, as HIV-1 and Influenza. Most methods for viral diversity estimation proposed so far are intended to take benefit of the longer reads produced by some NGS platforms in order to estimate a population of haplotypes which represent the diversity of the original population. Our goal here is to take advantage of distinct virtues of a certain NGS platform – the platform SOLiD™ (Life Technologies) – that has not received much attention due to the short length of its reads, which renders haplotype estimation very difficult. However, the platform SOLiD™ has a very low error rate and extremely deep coverage per site and our method is designed to take advantage of these characteristics.

**Results.** We propose to measure the populational genetic diversity through a family of multinomial probability distributions indexed by the sites of the virus genome, each one representing the populational distribution of the diversity per site. Moreover, the implementation of the method focuses on two main optimization strategies: a read mapping/alignment procedure that aims at the recovery of the maximum possible number of short-reads; the estimation of the multinomial parameters through a Bayesian approach based on Dirichlet distributions inspired by word count in text modeling. The Bayesian approach, unlike simple frequency counting, allows one to take into account the prior information of the control population within the inference of a posterior experimental condition and provides a natural way to separate signal from noise, since it automatically furnishes Bayesian confidence intervals.

**Conclusions.** The methods described in this paper have been implemented as an integrated tool called *Tanden* (Tool for Analysis of Diversity in Viral Populations) and tested on samples obtained from HIV-1 strain NL4-3 (group M, subtype B) cultivations on primary human cell cultures in many distinct viral propagation conditions. *Tanden* is written in C# (Microsoft), runs on the Windows operating system, and can be downloaded from: http://tanden.url.ph/

## Background

Viral biology is surprisingly diversified, and viruses with RNA genomes are recognized to generate particularly mutant-rich populations called *quasi-species*. The genetic heterogeneity characteristic of viral quasi-species is largely due to high mutational rates combined with an elevated populational size [1]. Human Immunodeficiency virus 1 (HIV-1), as an example, has a mean substitution rate of order $10^{-5}$ per nucleotide position [2]; that is by far higher than those of cellular organisms [3, 4] and assures a constant viral mutant production.

Next-Generation Sequencing (NGS) platforms have been used mainly for the *de novo* sequencing of viruses. However, more recently, new interest arose in re-sequencing known virus genomes using NGS to study the diversity of viral populations. All NGS platforms produce short segments of DNA, called *reads*, which provide only imperfect and incomplete information about the structure of the viral population. Sequencing errors and length of reads are factors that must be taken into account in the analysis of data obtained from NGS viral quasi-species. In addition, reverse transcription and PCR amplification are procedures prone to errors. The impact of these errors on studies of viral diversity could be huge (see below), therefore one wants to separate true genetic variation from methodological noise and if both are of the same order of magnitude the task  becomes virtually impossible.

Regarding the development of tools to estimate genetic diversity of viral populations, the most commonly used NGS platforms are the 454™ (Life Sciences/Roche), mainly due to their capacity to produce long reads but now in disuse, and the Illumina™ (Solexa). The ability to produce relatively long sequences favors the development of methods aiming at the *haplotype reconstruction* of the representative particles in population [5, 6]. However, the propagation of sequencing errors is a serious problem in these methods, requiring the development of procedures for error correction, which my introduce unwanted biases. In general, the fraction of wrong reads increases with the error rate and the average length. The expected proportion of reads with at least one sequencing error as a function of the error rate per base $\varepsilon$ and the average length $L$ of reads is given by $1-(1-\varepsilon)^L$ [7]. As the estimated error rate of 454™ is about 0.1% - 0.5% and Illumina™ error rates are in the range of 0.1% - 1% [8], with an average length of reads from 400 bp up to 1000 bp, the proportion of reads with at least one error is in the range 35% - 90%. The platform SOLiD™ (Life Technologies), for instance, is at the other end of the spectrum. With reads of short length, of at most 50 bases (the main limitation for the construction of haplotypes) and estimated error rate of 0.06% [8], the proportion of reads with at least one error is around 2%. Recently, a different solution to the problem of sequencing errors has been proposed [9], based on the development of high-fidelity sequencing protocols [10].

A more serious challenge concerns the *NP-hardness* of combinatorial optimization problem associated with the assembly of all possible haplotypes [11]. In fact, some approximate solution must be employed and a crucial hindering factor is the rate between the size of the reads and the size of the genomic region being reconstructed. For instance, it has been reported [9] that short read lengths (less than 100 base pairs) dramatically inhibit reconstruction of genomes of length 3400 base pairs, evidenced by the failing to produce any complete genome.

As mentioned before, the ability of the other NGS platforms to produce relatively long sequences have been a great stimulus to the development of methods for building *haplotype representatives* of the particles in population and the vast majority of softwares for viral diversity estimation that have been proposed until very recently adopt this perspective [5]. The aim of this work is to propose a different approach to genetic diversity evaluation that takes advantage of the low error rate and the high depth of coverage per site inherent to the SOLiD™ platform and therefore we shall considerably depart from the developments towards haplotype recosntruction. Indeed, although the short length of the reads produced by the SOLiD™ platform essentially hinders haplotype reconstruction, it is possible to measure the populational genetic diversity through probability distributions along the genome (one per site) and this approach is enhanced by the highly deep coverage provided by the SOLiD™ platform.

A recent study [12] comparatively assessed the performance of some NGS platforms (including 454™ and Illumina™) and reported an average (range) coverage of ~23000 reads (5000-47000) for the Illumina™ and ~7000 reads (2000-22000) for the 454$^{TM}$. We were able to achieve an average (range) coverage of ~50000 reads (10000-150000), for instance, see Figure 1. In addition, the low error rate of 0.06 % provided by the SOLiD$^{TM}$ platform virtually eliminates the necessity of any error correction procedure. Instead, we use the estimated probability distributions to separate signal from noise.

The first step in nucleotide sequence analysis is read mapping/alignment. This is important for many bioinformatics applications, as exemplified by nucleic acid conformational structure prediction and phylogeny studies [13, 14]. As expected, this is also an important aspect for Next Generation Sequencing (NGS) data analysis involving all the different platforms as Ion Torrent$^{TM}$ (Life Technologies), SOLiD$^{TM}$, 454$^{TM}$ and Illumina$^{TM}$ [15, 16] and others. Nowadays users can choose from a panoply of tools for mapping and indexing NGS reads, available on-line and for download. MAQ (Mapping and Assembly with Qualities) [17], BWA (Burrows-Wheeler Alignment Tool) [18], BFAST (Blat-like Fast Accurate Search Tool) [19], Bowtie [20] and MOSAIK [21] are examples of such alternatives. Those tools allow the fast mapping and alignment of reads belonging to genomes ranging from $10^6$ to $10^9$ bp in length. A common point between these tools is the use of scaffolding reference sequences [18, 22, 23].

After the read mapping is finished, the following step consists in the choice of a strategy for statistical inference. There is a wide variety of methods depending on the scope and the goals of the analysis: (i) consensus generation, (ii) *single nucleotide variant* (SNV), also called *single position diversity estimation*, (iii) *local diversity estimation* and (iv) read graph-based haplotype reconstruction, also known as *global diversity estimation*, see [7, 24] for a thoroughly explanation of these concepts. Existing tools for genetic diversity evaluation of viral NGS sequences, intended for 454$^{TM}$ and Illumina$^{TM}$ platforms [7, 24–31], are based on several techniques aiming at haplotype reconstruction [28, 32–37].

In order to estimate the populational diversity without resorting to haplotype reconstruction, we propose to measure the populational genetic diversity through a family of multinomial probability distributions indexed by the sites of the virus genome, each one representing the populational distribution of the diversity per site. Moreover, the estimation of the multinomial parameters is made through a Bayesian approach based on Dirichlet distributions inspired by word count in text modeling. The choice of a Bayesian estimation is crucial for our method, because it allows one to take into account the prior information of the control population within the inference of a posterior experimental condition by means of Bayes theorem and thus relates two temporally connected events. It also provides a natural way to separate signal from noise, since it automatically furnishes credible (or Bayesian confidence) intervals. On the other hand, simple frequency counting, as one could suggest, is neither appropriate nor sufficient for the implementation of the method. First, because it is not possible to include prior knowledge into the estimation (for example, we know in advance the genome of the control) and second, it does not provide a natural way to obtain confidence values for the estimate.

In summary, we sought to build an analysis platform suitable to address the problem of estimation of the populational diversity of RNA viruses. Due to high mutational rates and accelerated populational turnover RNA viruses constitute ensembles of variants. These ensembles known as *quasi-species* behave as a single and coherent organism and host pressures are thought to act on the ensemble of variants rather than on individual particles [38]. Based on the above assumptions and the fact that the SOLiD™ platform generates an extremely high number of reads allowing for a deep and extensive coverage of the data with very low error rate, we propose a simple approach – based on very few elementary assumptions – to the estimation of genetic diversity of viral populations.

Unfortunately, we could not find any other method or software in the literature, which uses a similar form to represent the viral population diversity as a family of distributions index by the genome – all other proposals are aimed at haplotype reconstruction and need longer reads (more than 100 bases). Any attempt to make comparison between such different aproaches would be misleading, therefore it is not our aim here to make the point if the method presented here is an improvement over (non-)existing similar ones. In fact, we believe that the approach proposed here should not be considered alternative or rival, but complementary, to haplotype reconstruction.

The technical analysis and conclusions about the experiments reported in this paper and used as test cases, are not necessary to the description of the methods introduced here and will be the subject of another publication [39].

## Implementation

Here we describe the main steps of our method. There are two stages, the first is the read mapping/alignment and the second is the nucleotide inference. The method presented here works, in principle, for data generated in any NGS machine, as long as the data is stored in the *FASTA file format*, even several outputs from different platforms (with distinct read lengths) may be combined and analyzed simultaneously.

### *1. Experimental procedure and preparation*

The computational tool developed here assumes that there is one, or even several, *experimental condition(s)*, representing different viral propagation situations, all of which, having the same viral population as the infecting source. After a determined number of replicative cycles an extracted sample from each experimental condition may be sequenced. A sample from the initial population prior to the infections must also be sequenced and will be referred as the *control experiment*. Raw data from the sequencing must be treated according to the standard procedures of the specific NGS platform [40] up to the generation of FASTA files, which are the standard type of input file adopted in our implementation. See Table 1 for a summary of data analyzed.

### *2. Read Mapping/Alignment*

The main goal of this step is the mapping of reads with 50 nucleotides or more originating from the NGS platform to a database of reference sequences. The database may contain several sequences, which must be aligned amongst themselves The read mapping is performed using a local executable of BLAST [41] with the default options. The criteria for retaining the reads are the following: (i) it must align at least 45 nucleotides and (ii) have the lowest *e*-value score. A first alignment attempt is made with sequences from reads in the forward sense; in case of no match, a second attempt with the reverse complementary sequence is performed. Moreover, since we are using several references, the output can, in principle, display the same number of matches as there is reference sequences. The criteria for the selection of the most suitable alignment option are the following (in this order): (i) the lowest *e*-value score and (ii) the lowest *Hamming distance* from the consensus sequence obtained from of the control. The alignment strategy described above is set as default, but some of its parameters can be changed according to some specific purposes or simply for increasing processing speed. Finally, it is possible to create suitable reference databases according to a specific research purposes.

*3. Nucleotide Estimation*

The probability distributions of the possible nucleotides (A,T,C,G) at each position of the genome are estimated from the aligned data. In this respect, our approach may be classified as a *diversity estimation in single positions*. The idea is that at each position in the genome the probability distribution is given by a *multinomial distribution*, determined by four probabilities ($p_A$, $p_T$, $p_C$, $p_G$) satisfying $p_A+p_T+p_C+p_G=1$. These probabilities represent the fraction of the population that has each of the four associated nucleotides at the corresponding site. Thus, one has a family of multinomial distributions indexed by the sites of the genome. In this sense one can view the nucleic acid sequences of a population as a collection of text documents and the task is to estimate the frequency of each character in each document.

The Bayesian approach provides an adequate framework [42, 43] for the inference of categorical data without the shortcomings of a simple frequency counting procedure. In this framework, the first step consists in using the *experimental control* as the input for estimation of a initial distribution. Then, in the second step, one considers this distribution as the *prior distribution* together with the sequenced data form the experimental conditions one uses *Bayes formula* to compute the *posterior distribution*. In the case of categorical data there is a canonical procedure based on the *conjugate* *prior distribution* associated with the multinomial distribution, called *Dirichlet distribution*.

A $n$-dimensional *Dirichlet distribution* is defined in by a smooth probability density function on the set Δ of $n$-dimensional multinomial distributions, which is parametrized as $\Delta_n = \{(p_1,\ldots,p_{n-1}) : p_1+\ldots+p_{n-1} \leq 1\}$, here $n$ is the number of distinct categories (states) that can observed and $p_k$ is the probability of observing the $k$-th category, for $k = 1,\ldots,n$ with $p_n = 1-p_1+\ldots+p_{n-1}$ . The *Dirichlet probability density* function is given by

$$\mathrm{Dir}(p_1,\ldots,p_{n-1} | \alpha_1,\ldots,\alpha_n) = 1/\mathrm{B}(\boldsymbol{\alpha}) \prod_k p_k^{\alpha_k-1} ,$$

where B(**α**) is a normalizing factor defined in terms of the *gamma function* Γ as

$$\mathrm{B}(\boldsymbol{\alpha}) = \prod_k \Gamma(\alpha_k) / \Gamma(\textstyle\sum_k \alpha_k).$$

for a vector $\boldsymbol{\alpha} = (\alpha_1,\ldots,\alpha_n)$. Note that the choice $(\alpha_1,\ldots,\alpha_n) = (1,\ldots,1)$ gives the uniform distribution (the *flat or uninformative prior*) on $\Delta_n$ with mass equal to the volume of $\Delta_n$ : $\mathrm{B}(1,\ldots,1) = 1/\Gamma(n) = 1/(n-1)!$. Thus a member in the family of Dirichlet distributions is characterized by a $n$-tuple of positive numbers $\boldsymbol{\alpha} = (\alpha_1,\ldots,\alpha_n)$ called *hyper-parameters* – however, unlike the multinomial parameters that must sum to one, the hyper-parameters are unconstrained. In our case, the Dirichlet distribution of each site is parametrized by the quadruple $(\alpha_A, \alpha_T, \alpha_C, \alpha_G)$. The hyper-parameters corresponding to each site of the genome must be estimated from the data.

The Dirichlet hyper-parameters associated to the control experiment can be obtained by maximum likelihood estimation (MLE) through the Newton-Raphson method. The *log-likelihood function* $g$ of the Dirichlet distribution is given by $g = N \log L$ and

$$\log L(\alpha_1,\ldots,\alpha_n | \bar{p}_1,\ldots,\bar{p}_n) = \log \Gamma(\textstyle\sum_k \alpha_k) - \sum_k \log \Gamma(\alpha_k) + \sum_k (\alpha_k-1) \log \bar{p}_k ,$$

where $N$ is the *sample size* and $\log \bar{p}_k = 1/N \sum_j \log p_{jk}$ ($j$=1,…,$N$, $k$=1,…,$n$) is called the *sufficient statistics* associated to a sample of $n$-categorical vector observations $\{\boldsymbol{p}_1,\ldots,\boldsymbol{p}_N\}$ of sample size $N$. Thus each vector $\boldsymbol{p}_j=(p_{j1},\ldots,p_{jn})$ has $n$ components, each component $p_{jk}$ is the frequency of the $k$-th category at the $j$-th sample.

The *Newton-Raphson method* for this log-likelihood function $g$ amounts to the iteration of the following fixed-point scheme [42]:

$$\boldsymbol{\alpha}^{\text{new}} = \boldsymbol{\alpha}^{\text{old}} + [H^{-1}\,\nabla g](\boldsymbol{\alpha}^{\text{old}}),$$

where $\boldsymbol{\alpha}^{\text{new}}$ and $\boldsymbol{\alpha}^{\text{old}}$ are vectors of Dirichlet hyper-parameters, $\nabla g(\boldsymbol{\alpha})$ is the *gradient vector* of the log-likelihood function $g$, with components

$$[\nabla g(\boldsymbol{\alpha})]_k = \Psi(\sum_k \alpha_k) - \Psi(\alpha_k) + \log \bar{p}_k\ ,$$

where $\Psi = (\log \Gamma)'$ is the *digamma function*. The matrix $H^{-1}(\boldsymbol{\alpha})$ is the inverse of the *hessian matrix* of the log-likelihood function $g$ and the product $(H^{-1}\ \nabla g)(\boldsymbol{\alpha})$ has components $[(H^{-1}\nabla g)(\boldsymbol{\alpha})]_k$ given by

$$[(H^{-1}\nabla g)(\boldsymbol{\alpha})]_k = (\Psi'(\alpha_k))^{-1}\ ((\nabla g)_k + (\sum_l (\nabla g)_l\ /\ \Psi'(\alpha_l))\ /\ (1/\Psi'(\sum_l \alpha_l) - \sum_l 1/\Psi'(\alpha_l))),$$

where $\Psi'$ is the *trigamma function* ($k$,$l$=1,…,$n$). Several suggestions for the initialization step of the iteration scheme described above have appeared in the literature [43–45]. The proposal of Ronning [45] is the most suitable for the modified iteration scheme adopted here.

Since we are dealing with a sparse estimation problem in the sense that one of the categories occur with much higher frequency that the other three, we shall employ the *smoothed sufficient statistics* defined by introducing a small parameter $\eta$ and setting $p_{jk} = M_{jk} / M$, where $M_{jk}$ is the number of occurrences of the $k$-th category at the $j$-th sample, $M$ is total number of observations at the $j$-th sample and $p_{jk} = \eta$ if there is no occurrence of the $k$-th category at the $j$-th sample. The *smoothing parameter* $\eta$ acts as “background noise” representing sequencing and PCR errors that can not be removed. However it can be suitably tuned in order to account for the true variability of the data. When this procedure is applied to the control (a “clonal” population) one would expect no diversity at all. However, that is not completely true and, in fact, even the control should display some variability (mainly due to sequencing errors). Since the expected error rate $\varepsilon$ is around $6\times10^{-4}$ a value of the order of $10^{-5}$ is a reasonable choice for $\eta$. In the smoothed version of the Newton-Raphson iteration scheme, Ronning's initialization step is given by setting $\boldsymbol{\alpha}^{\text{old}} = (\eta,\ldots,\eta)$.

The sufficient statistics is computed by a simple re-sampling procedure [46, 47] in order to generate sequences of categorical observations from the raw sequenced data, by randomly sampling nucleotides form each aligned position. Here, the imperfect clonality of the control is useful, since it ensure that the re-sampled ensemble has some variability, which is consistent with having a small non-zero smoothing parameter. The re-sampling procedure has one parameter that can be adjusted by the user: the relative size of observations given as a fraction $0 < z < 1$ of the whole set of nucleotides $C$ covering the given site. If the number of bases covering the given site is $C$ then $M=zC$ is the number of observations used to compute one sample vector $\boldsymbol{p}_j=(p_{j1},\ldots,p_{jn})$ and the corresponding sample size $N$ is given by (the integer part of) the logarithm of the total number of all possible sample vectors:

$$N = [\ \log\Gamma(C) - \log\Gamma(zC) - \log\Gamma((1-z)C)\ ]\ .$$

*Stirling's formula* gives the following approximation in terms of $C$:

$$N \approx C\,(-z\log z - (1-z)\log(1-z))/\log 2\ .$$

For instance, for the default value of $z$, which is 80%, one has a sample of size $N \approx 0.7C$, each sample vector computed from $0.8M$ nucleotides. On the other hand, the value $z$ = 50% gives a sample of size $N \approx C$, each sample vector computed from $0.5M$ nucleotides

Once the hyper-parameters of the prior distribution are estimated, they must be used together with the sequenced data of the other experimental conditions in order to compute the hyper-parameters of the posterior distributions by *Bayes formula*, which in the case of Dirichlet distributions can be computed explicitly (since it is a conjugate prior): if $(\alpha_1,\ldots,\alpha_n)$ is a vector of hyper-parameters of a Dirichlet prior distribution and the counts of each of the $k$ categories in an experiment are $(c_1,\ldots,c_n)$ then the posterior distribution is also a Dirichlet distribution with hyper-parameters $(\alpha_1+c_1,\ldots,\alpha_n+c_n)$. As a result, one obtains a family of Dirichlet probability distributions for every experimental condition, including the control experiment.

In order to obtain point estimates of categorical probabilities per site for each experimental condition ($p_A$, $p_T$, $p_C$, $p_G$), one may use a central tendency measure of the corresponding Dirichlet distribution (see [42]). Let $\boldsymbol{X} = (X_1,\ldots,X_n)$ be a random vector distributed according to a Dirichlet distribution with corresponding hyper-parameters $(\alpha_1,\ldots,\alpha_n)$ then the number $s = \alpha_1+\ldots+\alpha_n$ is called the *concentration parameter* of the corresponding Dirichlet distribution. It provides a measure of the "quality" of the inference: the greater the value of s the better is the "precision" of the inference (see [42]). The *expectation value* of $\boldsymbol{X}$ is

$$\mathbf{E}(X_k) = \alpha_k / s \, .$$

Confidence values associated to the point estimates may be defined in terms of a dispersion measure of the corresponding Dirichlet distribution. The *variance* of $\boldsymbol{X}$ is given by

$$\mathbf{Var}(X_k) = \alpha_k \, (s - \alpha_k) \, / \, (s^2 \, (s + 1)) \, .$$

Since the marginal distribution of each $X_k$ is a one-dimensional Dirichlet distribution, also known as *Beta distribution*, the *standard deviation of the mean* $\sigma(X)=\sqrt{\mathbf{Var}(X)}$ may be used to construct *Bayesian credible intervals* about the expectation value.

The *maximum a posteriori* (MAP) estimate, which is given by the *mode* of $\boldsymbol{X}$, has become a very popular method of point estimation [9]. Moreover, the coordinates $\bar{x}_k$ of the mode of $\boldsymbol{X}$ may be directly calculated in terms of the hyper-parameters when $\alpha_k > 1$ $(k = 1,\ldots,n)$:

$$\bar{x}_k = \alpha_k / (s - n) \, .$$

This is much simpler than the contrived expectation-maximization (EM) approximate schemes usually employed to obtain the MAP estimate from a log-likelihood function, in which case approximations are unavoidable, since this function is non-convex.

The the *standard deviation of the mean* $\sigma(X_k)$ may be used to define credible intervals about the mode as well. Since the Beta distribution is unimodal, when all $\alpha_k > 1$ ($k = 1,\ldots,n$), and has finite variance, a 3-sigma interval around the mean or the mode would provide about 95% of confidence in the prediction (this is a general consequence of the Gauss-Vysochanskij-Petunin inequality, see [48]).

Finally, we should note that the inference procedure explained above is clearly not restricted to the case of four nucleotides (A,T,C,G). It is trivial to modify it in order to account for insertions and deletions, or to work with codons and amino-acids.

*4. Selection Criteria and Error Filtering*

Once the inference has been completed it is desirable to filter the errors and extract some subset of the data – for instance, most conserved sites, most variable sites, etc. In order to do so we have implemented two selection criteria based on simple quantities: (i) complementary probability per site and (ii) variational distance per site.

The *complementary probability per site* is defined as $p_{comp} = 1-\max\{p_A, p_T, p_C, p_G\}$ and it depends only on the probability distribution of each site. It provides a measure of how much the distribution is concentrated in one state. For instance, if the complementary probability at a site is high it means that there was variation in the site prior to the experiment.

The *variational distance per site* is a positive number between 0 and 2 defined by $vd = |p_A-p'_A|+|p_T-p'_T|+|p_C-p'_C|+|p_G-p'_G|$, where ($p_A$, $p_T$, $p_C$, $p_G$) is the probability distribution per site in the control data and ($p'_A$, $p'_T$, $p'_C$, $p'_G$) is the probability distribution of the corresponding site in the experimental condition. It is a measure of the relative variation per site from the control to the experimental procedure data. If it is very low at a site it means that the site did not undergo significant changes in relation to the control.

The complementary probabilities and the variational distance can work as filters and the user must specify the thresholds for them. By using these two criteria in combination one may easily obtain some qualitative information about the behavior at a site.

## Results and Discussion

The method presented here was tested on samples obtained after the HIV-1 strain NL4-3 (group M, subtype B) cultivation on primary human cell cultures. Different viral propagation conditions were used – varying the cellular activation status, the co-receptor usage and the target cells. The pseudo-typed viruses produced in these experiments were able to perform exactly one round of the replicative cycle. As a whole, there were 7 experimental conditions in addition to the *control experiment* (Table 1).

### *1. Experimental procedure and preparation*

The experimental procedure was performed in accordance with the standard procedures of the NGS platform SOLiD™ [40], up to the generation of FASTA files, which are the input data of our computational tool. Standard Life Technologies guidelines were used during sample preparation and sequencing while using the SOLiD™ platform v. 3.0. The size of the FASTA files containing the reads of each condition is around 700Mb, consisting of about $10^7$ reads.

### *2. Read Mapping/Alignment*

The use of BLAST to perform the read mapping has two reasons: first the multiple reference sequences allowed by BLAST makes it advantageous for analysis of viral populations classically described as quasi-species, since we can include several variant genomes belonging to the same phylogenetic branch. Second, is the speed and the readiness of parallelization, since, for each experimental condition, we need to align about $10^7$ reads with 50 base pairs against $10^3$ reference sequences, each with $10^4$ sites, in the case of HIV. We used a reference database composed by 1258 sequences, properly aligned, of representatives of HIV-1 epidemics, most of them belonging to M group and its subtypes A, B, C, D, F, G, H, J, K, and recombinants. Some of the sequences are of complete HIV-1 genomes while others represented virtually (more than 90% complete) complete genomes. All sequences are available at NCBI/Genbank. The reference database also contains sequences from HIV-1 group O, SIVcpz and HIV-1 strain NL4-3 and is packaged together with the software.

Even though BLAST is capable of identifying alignments in both the forward and the reverse complementary senses, we have found that manually doing this significantly increases the retrieval of reads. Indeed, we have achieved an increase of 15% in the retrieval of reads. Since the procedure have been parallelized into multiple threads it has a minor effect on the performance when using a multi-core processor.

We were able to map around 90% of the reads, since the estimated fraction of reads with at least one error is around 2%, we have achieved an almost optimal retrieval of reads. For instance, Figure 1 shows the result of the alignment of the control experiment and the corresponding site coverage. The average site coverage is around 50000 reads with some peaks going beyond 150000 reads. The running-time on each experimental condition was around 30 hours on a Intel i7 (12 cores, clock of 3.30 GHz) with 32 GB of RAM memory and 2 TB of disk space. It is worthwhile mentioning that the program uses at most 3 cores and requires 2.8 GB of RAM memory to handle files with 700 MB, thus it is conceivable that the program could run on any computer matching this minimal configuration.

*3. Inference*

An important difficulty that should be overcome in order to implement the inference procedure for Dirichlet hyper-parameters in the context of nucleotides is due to the *sparsity*. Even with the high mutation rate displayed by viruses, there is a fair amount of nucleotide conservation. From a populational point of view, most of individuals will present the same nucleotide at a specific genomic position, and only the less representative subgroups, if any, will present one of the three remaining possibilities.

The standard Bayesian method outlined in most textbooks, where one usually chooses an uninformative (uniform) prior distribution is appropriate for the general task of multinomial estimation [49], but generally provides poor results when used for *sparse* multinomial distributions. This is primarily a consequence of the erroneous assumption that all categories should be considered as equally possible values for each site. Indeed, sparse multinomial distributions are characterized by the fact that only a few symbols actually occur (site conservation). In such cases, applying the standard method will give too much weight to symbols that never occur and consequently give a poor estimate of the true distribution. This issue becomes critical in our case when treating data obtained from the control experiment, which, in principle, is a *clonal population*, where one expects a uniquely well-defined nucleotide at each site and thus the Dirichlet likelihood function would be identically zero.

The sparsity problem is usually solved in the literature of text modeling by introducing a *smoothing parameter* $\eta$ and modifying the Newton-Raphson method in such a way that the sufficient statistics does not have any zero entry. In practice, this may result in an over-smoothed distribution, but one can choose a small enough value for $\eta$ in such a way that all the rare events do not have the same probability of appearing in all states [50].

*4. Validation*

The method described here contains some heuristic decisions that should be justified and properly validated. We have performed validation procedures at both stages of the method (alignment and inference), attaining very good concordance with the expected results.

In order to assess the reliability of the read mapping procedure and validate this stage, the *quality values* (QV) of the reads have been used as a proxy. The SOLiD$^{TM}$ platform outputs two files after primary analysis [43, 44]: a sequence file in color-space and a quality file containing the corresponding quality values. The QV of a read is a positive integer ranging from 0 to 50 and is given by the logarithm of the inverse probability of the color call being inaccurate, *i.e.* the higher the QV the higher the confidence in the color call's accuracy. By computing the *distribution of quality values* of the reads in each experimental condition and the control, prior and after the alignment, and comparing them, it is observed that they are almost identical (see Figure 2). This shows that the alignment procedure does not introduces any bias towards higher or lower quality values. The reliability of the read mapping procedure is guaranteed by the stringency of the criteria for retaining the reads.

The validation of the read mapping procedure was done by computing the distributions of all quality values of each condition prior and after the alignment (see Figure 2). The mean value of the QV distributions remained unchanged after alignment. Likewise, at both steps of the process more than 80% of reads had QV comprised between 20 and 32, assuring that the quality of the retrieved sequences was preserved and no bias was introduced.

The validation of the nucleotide inference step is performed at two points. The re-sampling procedure has been validated by comparing, at each site, the nucleotide frequencies obtained from all the reads that cover the site with the nucleotide frequencies obtained from the sampled reads that cover the site. It is observed that both frequencies agree with high precision (up to order $10^{-4}$). This ensures that the sufficient statistics obtained is the correct one. The validation of the implementation of the Newton-Raphson scheme for the Dirichlet maximum likelihood is performed by computing the MLE for a standard data-set, which is not sparse. The data set for pollen counts analyzed in Mosiman is often used for testing Dirichlet maximum likelihood implementations (see [42]). Since our implementation has a smoothing parameter, it is expected that the obtained values converge to the known values when the parameter approaches zero. It is indeed observed that this convergence occurs, with perfect agreement occurring above the order of magnitude of the smoothing parameter.

We have included in the software the appropriate options for the user to perform these validation procedures. In particular, it is possible to run the Dirichlet MLE on any data set (with 4 categories, in the present version) given as a list of multinomial observations.

*5. Error filtering*

The concentration parameter *s* of the control is a measure of the quality of the inference: when $s > 1$ the inference may be considered meaningful. Sites with $s \leq 1$ may be excluded from further analysis. Sites with low value of *s* may happen due poor coverage and total conservation (all reads with the same nucleotide at that position).

The complementary probability of an *ideal clonal population* would be identically zero. However, in the smoothed inferential scheme proposed here it is expected that $p_{comp} \approx \eta$. Furthermore, due to sequencing and PCR errors (and other events which may have occurred prior to cloning the initial genome), $p_{comp}$ may, in fact, display a broad distribution over the genome (see Figure 3). In any case, the complementary probability of the control may be considered as an average error rate per site and its distribution over the genome may be used to set a cut off value for separating the signal from the noise (everything below this value should be considered noise). It is expected that MODE($p_{comp}$) ≈ $\eta$, that is, the majority of sites will behave as in a clonal population and this indeed is the case (see Table 2). However, due to sequencing errors it expected to have a concentration of $p_{comp}$ near the error rate $\varepsilon = 6\times10^{-4}$. Since the distribution of $p_{comp}$ is extremely skewed with a long tail, the median is a better measure of centrality than the mean value. In fact, we have found that MEDIAN($p_{comp}$) ≈ $\varepsilon$ as expected (see Table 2).

The expectation MEAN($p_{comp}$), which is very sensitive to the long tail, is a reasonable choice of a cut off value for noise filtering, a more conservative choice would be MEAN($p_{comp}$)+SD-MEAN($p_{comp}$). The cut off value for $p_{comp}$ can be used to obtain a cut off value for the variational distance as CUT-OFF(*vd*) =2×CUT-OFF($p_{comp}$), since *vd* is a piece-wise linear function of the probabilities.

After the nucleotide probability distributions of the control was computed we have found 40 genomic positions with concentration parameter *s* less or equal than 1. These genomic positions correspond to portions of the genome where the coverage dropped substantially in comparison with the mean coverage (~$10^2$ reads). These sites were excluded from the remaining analysis. The expectation is MEAN($p_{comp}$) = 0.002 (that is, probabilities are considered significantly distinct if they differ by more than 0.02%). The conservative choice of cut off value is given by MEAN+SD-MEAN = 0.002+0.018 = 0.02 (see Table 2). The complementary probability may also be used to make sure that the variability observed in the experimental conditions is not a feature that has been transferred from the control to the experimental condition. The distribution of the complementary probabilities of the control shows that 98.5% of genomic positions have $p_{comp} < 0.02$ (this means less than 2% of nucleotide variation). The remaining 1.5% genomic positions correspond to sites were the population acquired its variation prior to exposition to the experimental condition (see Figure 5).

The posterior probability distributions of all 7 experimental conditions were computed. Figure 5 presents the values of the variational distance between the control and one of the experimental conditions and its distribution is shown in Figure 5, lower panel. Considering the same conservative cut off value of 0.04 for the variational distance (Figure 5, upper panel), one has that 98% of the genomic positions felt under this threshold, these are sites that did not display nucleotide variation after exposition to the experimental condition. The remaining 2% genomic positions contains all the populational variation acquired after exactly one round of the replicative cycle. Further analysis of these results and their implications for the study of HIV will be subject of another publication [39].

**Conclusions**

High throughput sequencing technologies are constantly evolving and new platforms and refinements in the chemistry and base calling algorithms are constantly improving. Recently the PacBio™ sequencer has been gaining space as it produces long reads, but with a large number of randomly generated sequencing errors [51]. New approaches to sequencing using known technologies have been proposed, such as circle sequencing for Illumina [52]. We expect that the proposed approach, with slight modifications can be adopted for other technologies such as Ion Torrent™, Illumina™ (HiSeq, MiSeq and NextSeq) and PacBio™ .

We have described a platform suitable to address the problem of estimation of populational diversity of RNA viruses. Based on the fact that the SOLiD™ sequencing platforms generate an extremely high number of reads allowing for a deep and extensive coverage of the data with very low error rate, we propose to measure the populational genetic diversity through a family of probability distributions indexed by the sites of the genome, each one representing the populational distribution of the diversity. This approach allowed us to avoid some hard problems related to haplotype reconstruction and emphasize the main features of the sequencing technology used in this work, the SOLiD™ platform.

We have tested the method proposed here on samples obtained after the HIV-1 strain NL4-3 (group M, subtype B) cultivation on primary human cell cultures in many distinct viral propagation conditions, thus successfully demonstrating the capability of the method in handling large data-sets and delivering very clean results, suggesting that the software is a valuable tool for investigating the genetic diversity in viral populations. We have successfully demonstrated *Tanden*'s capability of handling large data-sets and delivering very clean results, suggesting that the software is a valuable tool for investigating the genetic diversity in viral populations as a complementary to some haplotype reconstruction method.

## Availability and requirements

**Project name:** Tanden

**Project web site:** http://tanden.url.ph/

**Operating systems:** Windows

**Programming language:** Microsoft - C#

**License:** free to all users under the LGPL license

**Minimum requirements:** 4GB RAM (16GB recommended), 500 GB disk space

## Competing interests

The authors declare that they have no competing interests.

**Authors' contributions**

JPZ: contributed to the statistical analysis, developed and implemented the software, drafted the manuscript; SNB: carried out the biological analysis, contributed to the statistical analysis, drafted the manuscript; ACV: contributed to NGS and bioinformatics analysis, drafted the manuscript; GCO: contributed to NGS bioinformatics analysis, drafted the manuscript; LMRJ: conceived the statistical analysis, contributed to NGS and biological analysis, drafted the manuscript; FA: conceived the statistical analysis, developed and implemented the software, contributed to NGS and biological analysis, drafted the manuscript. All authors read and approved the final manuscript before submission.

**Acknowledgements**

The ideas presented in this paper were developed in collaboration with Francisco A. R. Bosco, who suddenly passed away in December of 2012.

**Funding**

FAR Bosco received support from FAPESP (Brazil) through the grant 08/04531-2 (2009-2011). SNB received support from CAPES (Brazil) and FAPESP (Brazil). JPZ is supported by CAPES (Brazil). ACV and GCO thank the Program for Technological Development in Tools for Health PDTIS FIOCRUZ (RPT01F-NGS) for use of its facilities. LMRJ received support from the FAPESP project "In vitro study of the mutation rate of human immunodeficiency virus type 1 (HIV-1) in a single replication round", grant 2009/14543-0. FA is supported by CNPq (Brazil) through the grant PQ-306362/2012-0.

## Tables

| EXPERIMENTAL CONDITION | READS | NUCL. | MAPPED | TIME(HOUR) |
|---|---|---|---|---|
| HIV1 - Non-stimulated CD4/R5 | $1.11x10^7$ | $5.53x10^8$ | 91.81 % | 34.88 |
| HIV2 - Stimulated CD4/R5 | $1.04x10^7$ | $5.19x10^8$ | 90.84 % | 34.18 |
| HIV3 - Non-stimulated CD4/X4 | $1.10x10^7$ | $5.50x10^8$ | 91.22 % | 29.05 |
| HIV4 - Stimulated PBMC/X4 | $9.41x10^6$ | $4.71x10^8$ | 91.14 % | 13.63 |
| HIV5 - Stimulated PBMC/R5 | $1.14x10^7$ | $5.73x10^8$ | 90.49 % | 24.09 |
| HIV6 - Non-stimulated PBMC/X4 | $1.12x10^7$ | $5.62x10^8$ | 91.42 % | 27.54 |
| HIV7 - Non-stimulated PBMC/R5 | $1.19x10^7$ | $5.94x10^8$ | 93.27 % | 15.34 |
| HIV8 - Control (pNL4-3kfs) | $1.01x10^7$ | $5.05x10^8$ | 93.10 % | 30.14 |
| TOTAL | $8.65x10^7$ | $4.33x10^9$ | 91.67 % | 208.85 |

**Table 1:** Summary of data analyzed. The second column (READS) displays the number of reads sequenced in each condition, the third column (NUCL.) displays the number of nucleotides in each condition, the fourth column (MAPPED) displays the percentage of reads that have been mapped and the fifth column (TIME) displays the time elapsed in each mapping procedure.

| STATISTICS | $p_{comp}$ | $s$ |
|---|---|---|
| Mean | 0.00260 | 3.43 |
| Deviation | 0.01824 | 0.50 |
| Median | 0.00066 | 3.60 |
| Mode | 0.00001 | 3.74 |
| Minimum | 0.00001 | 1.01 |
| Maximum | 0.49242 | 6.26 |

**Table 2:** Summary statistics of the complementary probability ($p_{comp}$) and the concentration parameter ($s$) of the control, after removal of the genomic positions with concentration below 1.

## Figures

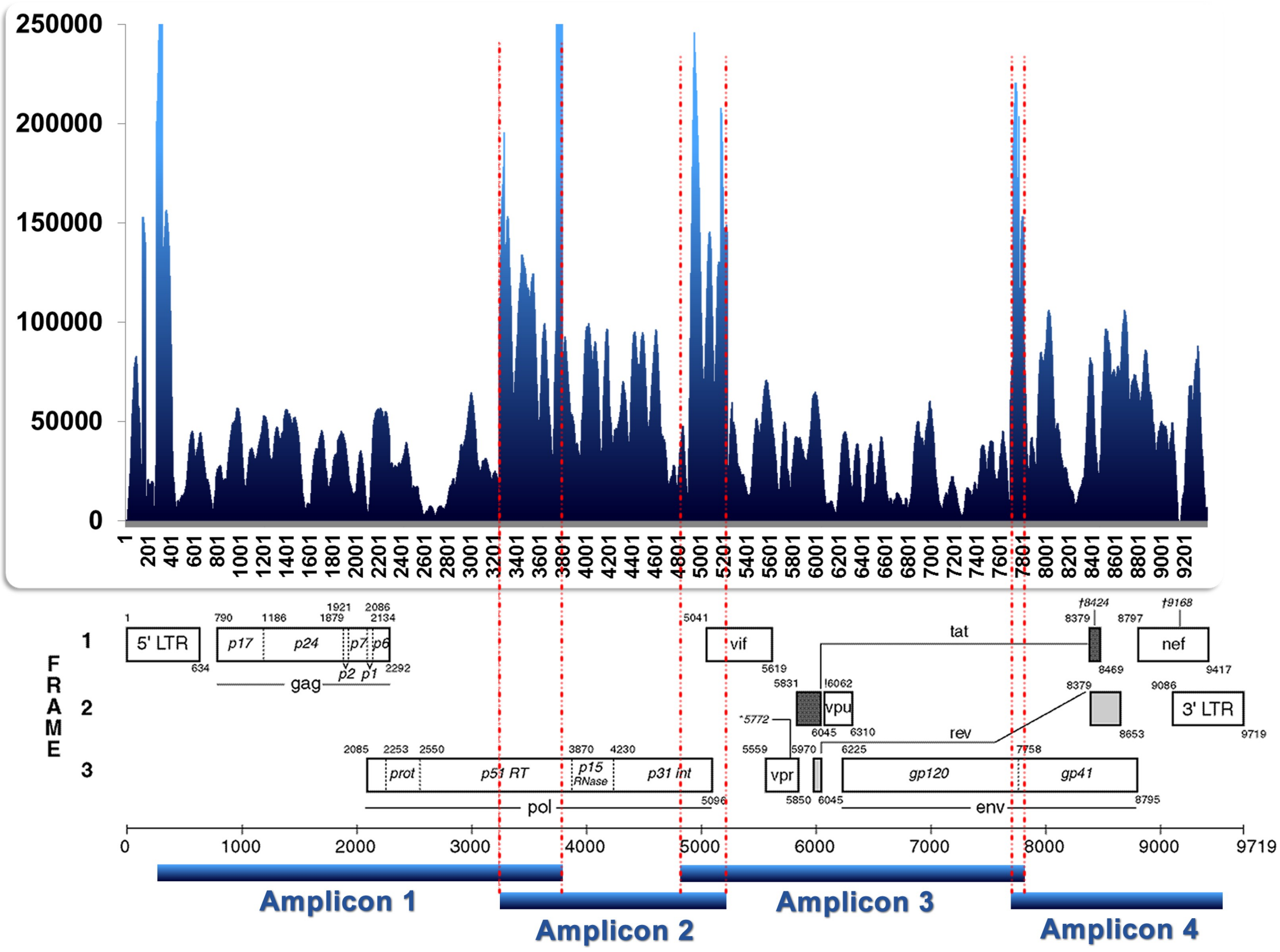

**Figure 1.** Depth and coverage of one SOLiD™ sequencing of the HIV-1 genome. The major peaks in the middle representing the most deeply covered regions coincide with the overlapping primers from the PCR step, an evidence that there is in fact some influence of pre-sequencing phases on the frequency of the short-reads retained in the alignment. The major peak in the beginning is related to the difficulties in mapping the LTR region. Other significant peaks maybe due to PCR artifacts, as well.

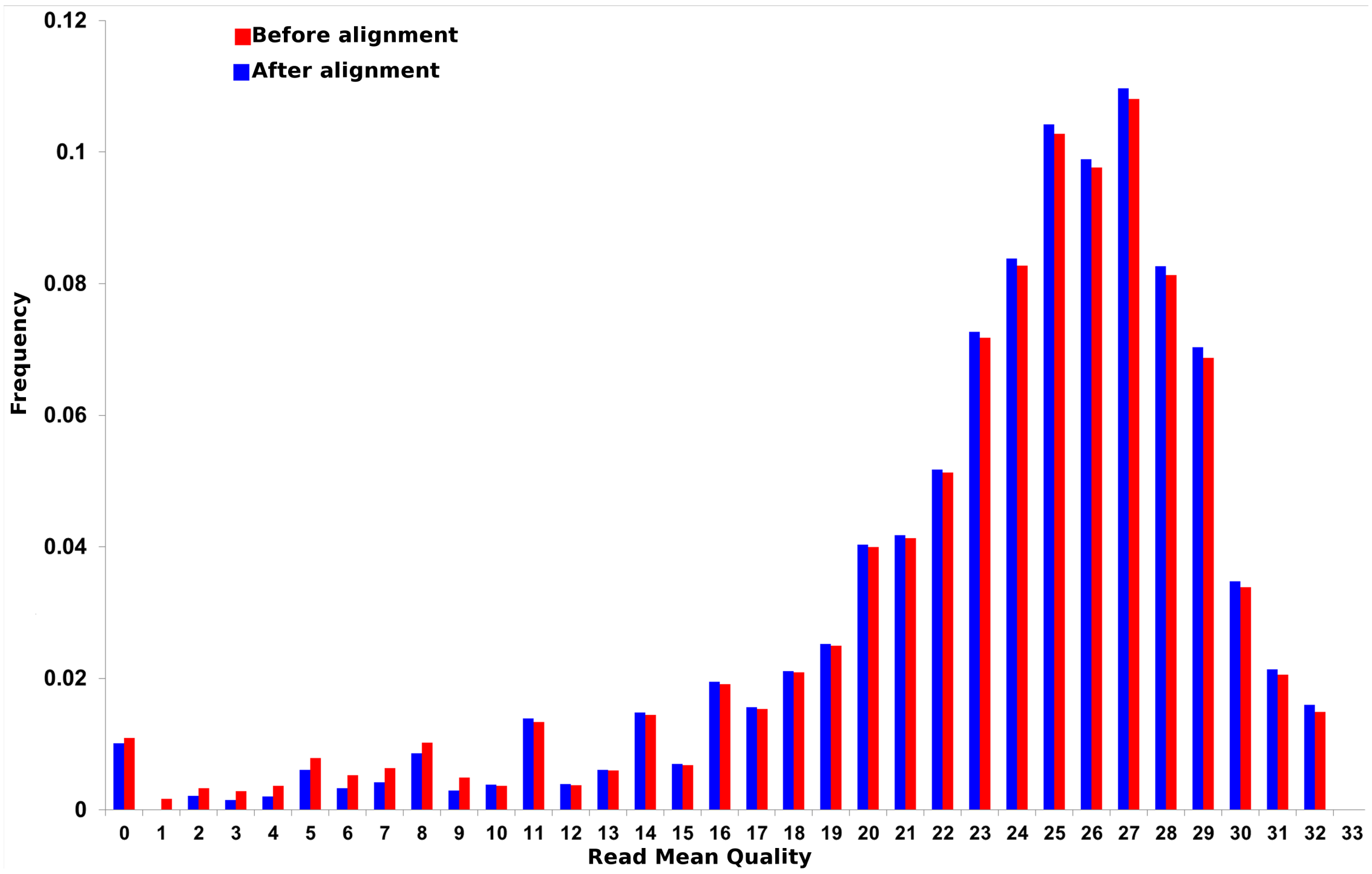

**Figure 2.** Quality values (QV) of the short-reads before (Blue) and after (Red) the alignment. The histogram contains the quality values on the horizontal axis and the proportion of short-reads in the vertical axis, displaying a large concentration of values around the average (approximately 23) and the majority (more than 90%) of short-reads in the range 16-32. Short-reads with quality values in this range are considered to have excellent fidelity.

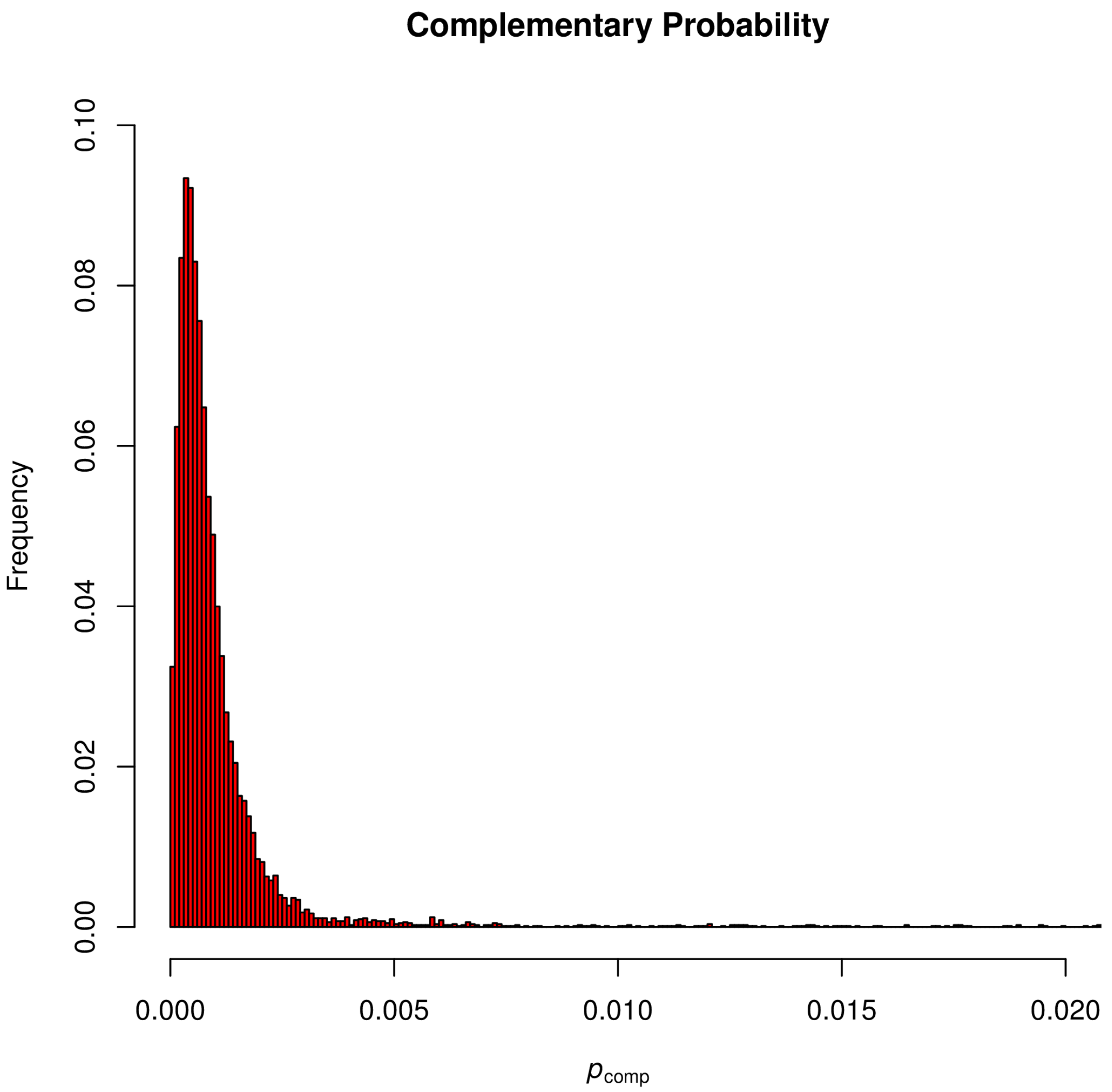

**Figure 3.** Histogram of the complementary probabilities of the control data. The complementary probability per site is defined as $p_{comp} = 1 - \max\{p_A, p_T, p_C, p_G\}$ and it depends only on the probability distribution of each site. The horizontal axis shows the values of complementary probabilities and the vertical axis the proportions of sites. The histogram contains the sites with $p_{comp} < 0.02$, which comprises 98.5% of all genome. These are the sites that have a unique dominant nucleotide with probability greater or equal than 0.98. The remaining 1.5% sites are the ones displaying some variability in the distribution of nucleotides.

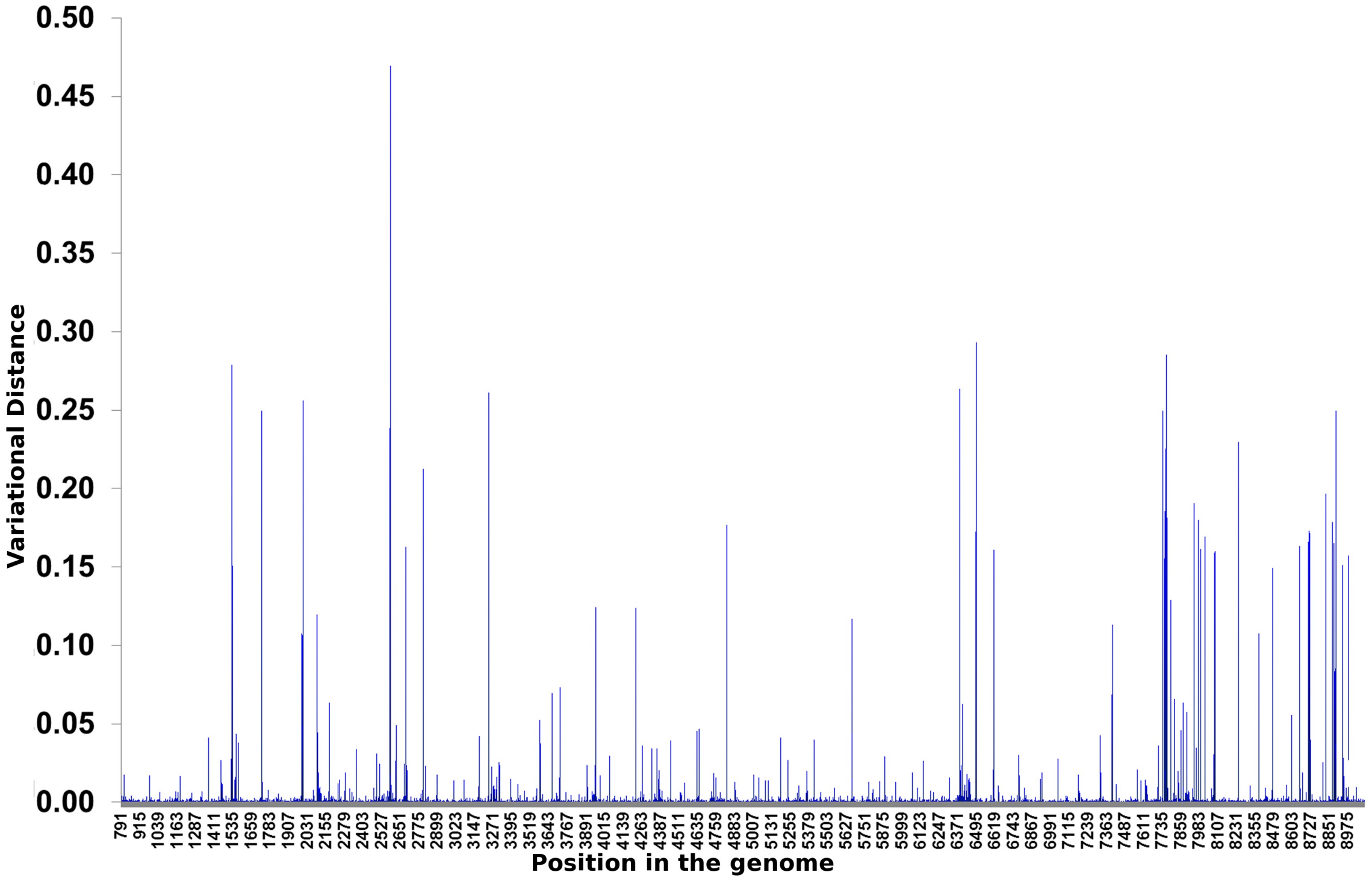

**Figure 4.** Variational distance (*vd*) between the control data and an experimental condition along the genome. The variational distance per site is defined by $vd = |p_A - p'_A| + |p_T - p'_T| + |p_C - p'_C| + |p_G - p'_G|$, where ($p_A$, $p_T$, $p_C$, $p_G$) is the probability distribution per site in the control data and ($p'_A$, $p'_T$, $p'_C$, $p'_G$) is the probability distribution of the corresponding site in the experimental condition. The horizontal axis shows the sites of the genome (with the LTR regions removed) and the vertical axis shows the corresponding variational distances. Applying the conservative cut-off value of 0.04 for *vd* one obtains the sites with significant variation.

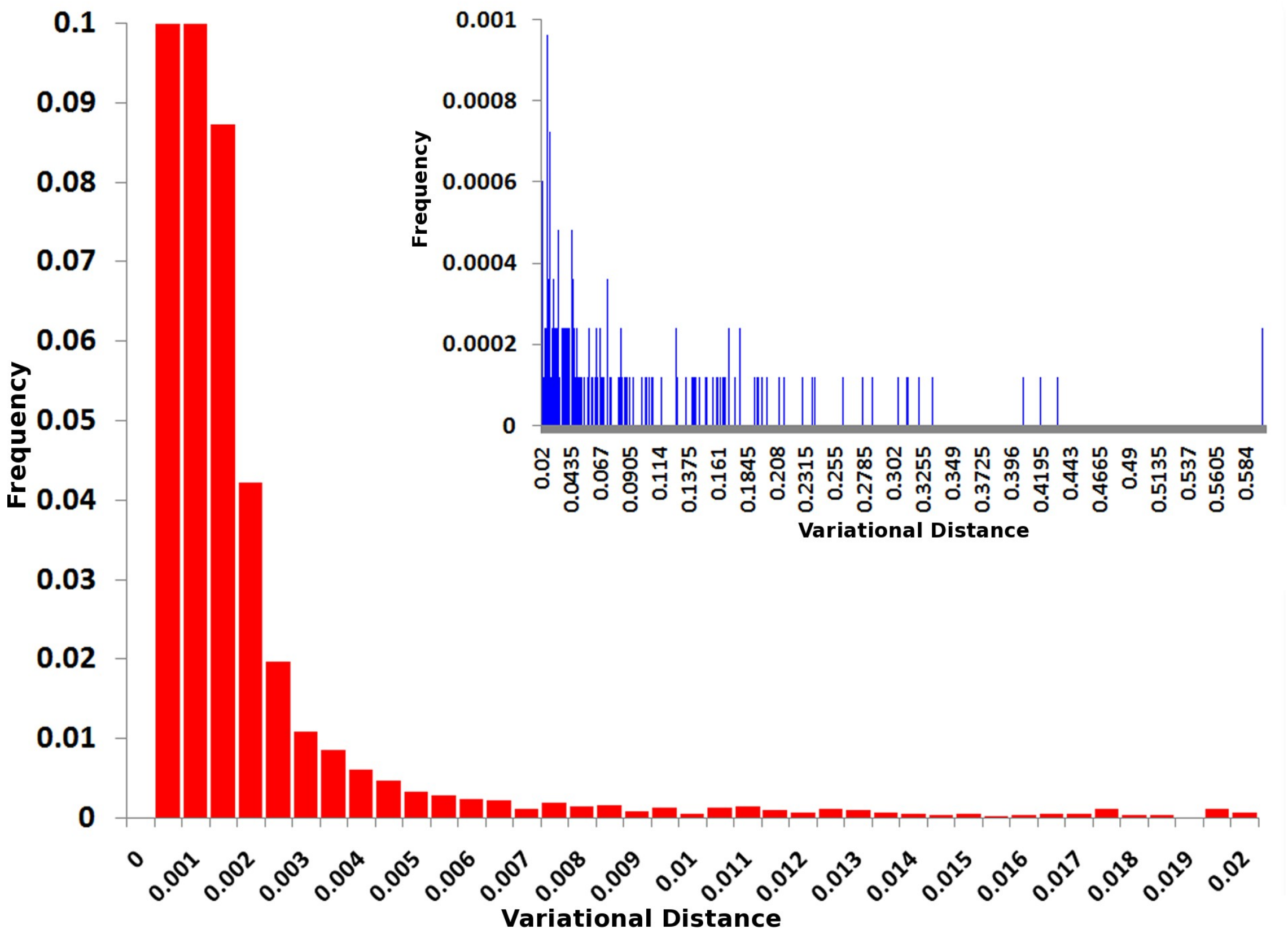

**Figure 5.** Histograms of the relative frequencies of the variational distances (*vd*). The horizontal axis shows the variational distances and the vertical axis shows the proportions of sites. The main panel (Red) shows the sites with *vd* < 0.04, which comprises 98% of all sites and the upper panel (Blue) shows the sites with *vd* > 0.04, which comprises 2% of all sites.